# Tailoring Fe/Ag Superparamagnetic Composites by Multilayer Deposition

J. Balogh, D. Kaptás, L. F. Kiss, and T. Pusztai

Research Institute for Solid State Physics and Optics, H-1525 Budapest, P. O. Box 49, Hungary

E. Szilágyi and Á. Tunyogi

KFKI Research Institute for Particle and Nuclear Physics, H-1525 Budapest, P. O. Box 49, Hungary

J. Swerts, S. Vandezande, K. Temst, C. Van Haesendonck

Laboratorium voor Vaste-Stoffysica en Magnetisme, K.U.Leuven, Celestijnenlaan 200 D, B-3001 Leuven, Belgium

The magnetic properties of Fe/Ag granular multilayers were examined by SQUID magnetization and Mössbauer spectroscopy measurements. Very thin (0.2 nm) discontinuous Fe layers show superparamagnetic properties that can be tailored by the thickness of both the magnetic and the spacer layers. The role of magnetic interactions was studied in novel heterostructures of superparamagnetic and ferromagnetic layers and the specific contribution of the ferromagnetic layers to the low field magnetic susceptibility was identified.

Control over the shape, size and spacing of the magnetic elements in nanoscale composites is an important issue, not only for applications, but also for our basic understanding of the interactions among magnetic elements. It has already been shown[1,2] that thin films of granular alloys can be prepared by alternate deposition of the constituents in the form of discontinuous multilayers or, as these are frequently called, granular multilayers. When compared to the co-deposition methods this has the advantage that the size and the distance of the clusters can be varied independently by the nominal thickness of the magnetic and the spacer layers, respectively, and self-organization processes[2] can contribute to the nanostructure formation. Fe-Ag co-deposited granular alloys[3,4,5] and multilayer structures[6,7,8,9] were investigated before, but granular multilayers have not yet been examined systematically.

In this article we will demonstrate two interesting properties of Fe-Ag granular multilayers:

(i) The magnetic grain size can be reduced not only by decreasing the thickness of the magnetic layers ($d_{Fe}$), but also by increasing the Ag layer thickness ($d_{Ag}$).

(ii) Continuous magnetic layers can be inserted into the granular multilayer structure without modifying the average magnetic grain size (derived from the superparamagnetic blocking temperature ($T_B$)) and the almost perpendicular magnetic alignment of the granular layers.

These properties can be exploited in tailoring the magnetic behavior of granular multilayers and in preparing novel magnetic composites with varying amount and sequence of the discontinuous and continuous layers. Regularly structured heterogeneous systems show interesting magnetoresistance[10,11] and magnetic behavior[12], but comparison of the heterostructures to granular alloys with random distribution of the magnetic elements will be beneficial, as well. Samples consisting of superparamagnetic (SPM) and ferromagnetic (FM) layers stacked in different order are very well suited to study the role of interactions between the different layers.

The magnetic properties were studied by a superconducting quantum interference device (SQUID) magnetometer and by $^{57}$Fe Mössbauer spectroscopy. The Fe/Ag multilayer samples were prepared by evaporating $^{57}$Fe and Ag from Knudsen cells with approximately 0.004 and 0.03 nm/sec evaporation rates, respectively, in a vacuum of $10^{-8}$ Pa onto rotating Si(111) substrates at room temperature. The nominal thickness was obtained from X-ray reflectivity analysis of the finite thickness fringes appearing in calibration films of the two separate materials. The evaporation always started with a Ag layer and was finished by a 5 nm thick Ag capping layer. The multilayer structure was examined by X-ray reflectivity measurements. Two samples containing only granular Fe layers, as will be seen from the magnetic properties, showed no peaks in the reflectivity measurements:

A)   [0.2 nm $^{57}$Fe / 5.4 nm Ag]$_{75}$ ,
B)   [0.2 nm $^{57}$Fe / 2.6 nm Ag]$_{75}$ .

On the other hand the multilayers containing continuous Fe layers of 1.5 nm nominal thickness in interleaved and separated form (see inset of Fig. 1):

C)   [(0.2 nm $^{57}$Fe / 2.6 nm Ag)$_3$ / (1.5 nm $^{57}$Fe / 2.6 nm Ag)]$_{25}$ ,
D)   [1.5 nm $^{57}$Fe / 2.6 nm Ag]$_{25}$ / [0.2 nm $^{57}$Fe / 2.6 nm Ag]$_{75}$

exhibit multilayer reflectivity peaks as illustrated in the main part of Fig. 1, although the finite thickness fringes are not resolved. The peak positions in Fig. 1 indicate that the actual periodicity is at least 20 % lower than the nominal one. The average Fe and Ag content of the samples was checked by backscattering spectrometry. The overall Fe concentrations of the Fe$_x$Ag$_{100-x}$ samples (x = 4, 12, 22 and 25 at%) agreed rather well with the nominal values (x = 5, 10, 22 and 22 at%) within the experimental error, but the total amount of $^{57}$Fe and Ag in the samples was 30 to 35 % less, indicating that both d$_{Fe}$ and d$_{Ag}$ are lower than the nominal values. In the following we will, however, use the nominal layer thickness.

    The temperature dependence of the bulk magnetization measured by the SQUID in an applied field of 1 mT after zero field cooling (ZFC) and field cooling (FC) in 1 mT is shown in Fig. 2. Samples A and B show magnetic irreversibility typical of SPM small particle systems[13] with a T$_B$ of 12 K and 40 K for samples A and B, respectively. In case of non-interacting SPM particles the magnetization is described by the Langevin function and T$_B$ is proportional to the volume of the magnetic clusters. For samples A and B the temperature and applied magnetic field dependence of the magnetization could be well fitted by a single Langevin function above T$_B$ and the number of Fe atoms in the SPM clusters was calculated to be 90 and 267 for samples A and B, respectively. A characteristic property of non-interacting SPM particles is the scaling of the magnetization curves measured at different temperatures when they are plotted as a function of H/T, i.e., the applied field divided by temperature. The right panels in Fig. 2 show that this scaling behavior can indeed be observed above T$_B$ for both samples. Small deviations between the curves measured above T$_B$ can be attributed to interaction between the magnetic particles and the observed deviations are comparable to those measured in co-deposited samples[5] with similar concentration of the magnetic particles. Since the effect of interactions on the blocking transition is not well understood[13], we will discuss below in more detail whether the observed increase of T$_B$ for smaller d$_{Ag}$ can be a result of interaction between the SPM layers or is rather due to a larger magnetic cluster size in spite of the same nominal d$_{Fe}$. A decreasing blocking temperature with increasing d$_{Ag}$ was also suggested for a series of samples with 0.7 nm nominal d$_{Fe}$ based on the observed[9] temperature dependence of the hyperfine field distributions. However, since the blocking temperatures were above 300 K, the blocking process could not be measured directly. Determining the size of the Fe clusters is very difficult in Fe/Ag nanostructures. Direct transmission electron microscopy investigations are restricted to samples on special substrate material[3] because sample preparation can easily modify the structure, while diffraction methods

suffer from the problem of the strong overlap between the diffraction lines of bcc Fe and fcc Ag. Therefore, we investigated whether magnetic interactions across the spacer layer play a role in the variation of $T_B$. If magnetic interactions between the layers are responsible for the higher $T_B$ of sample B, interleaving FM layers with similar spacing is expected to further increase $T_B$ and the samples with equal number of FM layers (C and D) will exhibit different magnetic properties in the interleaved (C) and the separated (D) form.

In the FC and ZFC magnetization curves of samples C and D (Fig. 2), small bumps around 40K can be observed in both samples, but the overall magnetic behavior is very different. As compared to sample B, the irreversibility temperature is shifted to higher but different temperatures for the two samples, however, taking into account the much larger magnetization values, this should be rather linked to the continuous Fe layers containing more than 70 % of the Fe atoms. The dominant contribution of a FM component is also reflected in the field dependence of the magnetization, as illustrated in the right panels of Fig. 2. The measured curves could be fitted above 0.2T external field as the sum of a constant (75% of the saturation magnetization value for C and 72 % for D) and a Langevin function due to a FM and a SPM fraction, respectively. The average number of Fe atoms in the SPM clusters, as calculated from the fitted paramagnetic moment and the saturation magnetization values, are quite similar: 258 for C and 232 for D. Although the behavior of the samples in high magnetic field is consistent with the number of FM and SPM layers in the samples, the origin of the irreversibility observed in small magnetic field remains to be explained.

The Mössbauer spectra of samples B, C, and D at 12 K and at room temperature are shown in Fig. 3 (a). Sample B is fully paramagnetic at room temperature and the average isomer shift (IS = 0.17 mm/s relative to α-Fe) and quadrupole splitting (QS = 0.42 mm/s) are comparable to those measured[14] on co-evaporated samples. The contribution of the granular Fe layers to the Mössbauer spectra can be clearly distinguished in samples C and D at room temperature, as indicated by the dotted line in Fig. 3 (a). While the granular layers are paramagnetic at room temperature, the 1.5 nm Fe layers show well-resolved hyperfine splitting in both samples. The relative number of Fe atoms in the 1.5 nm thick layers is well reflected by the spectral area under the magnetically split components: 76 (±2)% and 72 (±2)% for the interleaved and the separated samples, respectively. At low temperature a broad hyperfine field distribution can be observed in all the samples. In Mössbauer spectroscopy $T_B$ is generally defined[13] as the temperature where the area under the magnetically split component is equal to that under the paramagnetic one. We followed an approach that is model-independent and more effective when the sample is heterogeneous with a small SPM fraction and monitored the counts in a narrow velocity range around the paramagnetic peaks. The following quantity was evaluated:

$$\Sigma(T) = \sum_{peak} C(v,T) \bigg/ \sum_{base} C(v,T),$$

where $\sum_{peak}$ and $\sum_{base}$ run over velocity ranges around the paramagnetic peaks and at the baseline, respectively, and $C(v, T)$ is the number of counts measured at velocity $v$ and temperature $T$. The relevant velocity ranges are indicated in Fig. 3 (a). $\Sigma(T)$ is an S-shaped curve and $T_B$ can be defined as its inflection point. Figure 3 (b) shows $\Delta\Sigma(T)/\Delta T$ calculated numerically for samples B, C, and D. The position of the peak, i.e. the average $T_B$, agrees well for the three samples and can be connected to the small bumps in the ZFC magnetization curves. The slight smearing out of the transition temperature in sample C is consistent with the magnetization measurement, but no relaxation feature (significant change of $\Sigma(T)$ or broadening of the magnetically split lines) can be connected to the irreversibility of the bulk magnetization at higher temperatures.

At low temperatures the hyperfine field distributions for the Fe atoms in the SPM and in the FM layers strongly overlap, but the variation of the average line intensities also clearly reflects the blocking transition. The intensity of the six lines of a sextet is distributed as 3: $I_{2-5}$:1:1: $I_{2-5}$:3, where $I_{2-5} = 4\sin^2\alpha / (1 + \cos^2\alpha)$ is the intensity of the $\Delta m = 0$ transitions, and $\alpha$ is the angle between the direction of the γ-ray and the magnetic moment. The assymetry of the spectra in Fig. 3 (a) is due to correlations among the hyperfine field, isomer shift and quadrupole splitting parameters of the distributed sextets. The small relative intensity, $I_{2-5} = 0.6$ for sample B (these lines are marked by arrows in Fig. 3 (a).), indicates a close-to-perpendicular alignment of the magnetic moments. The magnetic components in the 300 K spectra of samples C and D could be fitted with $I_{2-5} = 3.5$ and 3.6, implying that the magnetization of the continuous layers lies almost in the sample plane. The temperature dependence of the average $I_{2-5}$ for samples C and D is shown in Fig. 3 (c). Around $T_B$ a decrease of $I_{2-5}$ occurs for both samples and the values at 12 K ($I_{2-5}$=2.7 and 2.5) are close to the weighted average of $I_{2-5}$= 0.6 for the granular layers and $I_{2-5}$ = 3.5 and 3.6 for the continuous layers taking into account the number of Fe atoms in the different layers. This suggests that both the perpendicular alignment of the granular layers and the in-plane alignment of the continuous layers are preserved in the interleaved sample.

From the Mössbauer measurements we can conclude that the average $T_B$ and the nearly perpendicular magnetic alignment of the granular layers remain intact in the interleaved sample. On one hand this indicates that the range of interactions affecting $T_B$ is shorter than the actual spacer layer thickness and implies that the observed decrease of $T_B$ with increasing $d_{Ag}$ results from a decrease of the average magnetic grain size. On the other hand the Mössbauer results also show that the large irreversibility of the low field magnetization observed in the heterostructures is not due to the granular layers. A similar effect was observed[5] in single continuous Fe layers capped with Ag and the irreversibility tempereture was rather insensitive to the Fe layer thickness. We propose to explain this kind of irreversibility in terms of domain wall motion in the FM layers. The largely different small field behavior of the heterostructures deviating only in the layer sequence supports this idea and points to the possible role of domain wall pinning at the interface. Such an effect was proposed[12] to be effective in the separated geometry, but the temperature dependence was not studied. This may as well play a role in the magnetic behavior of granular alloys when the concentration of the magnetic component is around the percolation threshold. The smearing out of the peak of the ZFC curve and the increase of the temperature where irreversibility sets in are generally attributed[5] to inter-particle interactions. Our results clearly demonstrate that thin FM regions of the samples can also account for the irreversibility.

In conclusion the possibility of tailoring nanoscale composites of superparamagnetic and ferromagnetic components by multilayer deposition was demonstrated for the Fe/Ag system. Mössbauer spectroscopy was applied to identify the specific contribution of the granular layers to the bulk magnetization and it was shown that the continuous magnetic layers inserted into the granular multilayer structure do not modify the average magnetic grain size and the almost perpendicular magnetic alignment of the granular layers. The possibility of control over the superparamagnetic grain size, the amount and the stacking sequence of the ferromagnetic fraction makes these heterostructures suitable model materials for understanding the magnetic behavior of granular composites with randomly distributed magnetic elements. In the present study we have found that varying the stacking sequence affected the low field magnetic susceptibility of the ferromagnetic layers more significantly than that of the superparamagnetic ones.

**Acknowledgements**

The authors acknowledge the financial support from the Hungarian Research Funds OTKA T034602, T038383, and T 046238 as well as from the Fund for Scientific Research - Flanders (FWO) and the Belgian IAP and Flemish GOA research programs. KT is a Postdoctoral Research Fellow of the FWO.

**References:**

[1] A. B. Pakhomov, B. K. Roberts, and K. M. Krishnan, Appl. Phys. Lett. 83, 4357 (2003).
[2] D. Babonneau, F. Petroff, J.-L. Maurice, F. Fettar, A. Vaurès, and A. Naudon, Appl. Phys. Lett. 76, 2892 (2000).
[3] G.-F. Hohl, T. Hihara, M. Sakurai, T. J. Konno, K. Sumiyama, F. Hensel, and K. Suzuki, Appl. Phys. Lett. 66, 385 (1995).
[4] C. L. Chien, John Q. Xiao, and J. Samuel Jiang, J. Appl. Phys. 73, 5309 (1993).
[5] C. Binns, M. J. Maher, Q. A. Pankhurst, D. Kechrakos, and K. N. Trohidou, Phys. Rev. B 66, 184413 (2002).
[6] P. J. Schurer, Z. Celinski, and B. Heinrich, Phys. Rev. B 51, 2506 (1995).
[7] G. Gladyszewski et al. Thin Solid Films 366, 51 (2000)
[8] R. Gupta, M. Weisheit, H. U. Krebs, and P. Schaaf, Phys. Rev. B 67, 075402 (2003).
[9] J. Balogh, D. Kaptás, T. Kemény, L. F. Kiss, T. Pusztai, and I. Vincze, Hyperfine Interactions 141/142, 13–20 (2002).
[10] D. Bozec, M. A. Howson, B. J. Hickey, S. Shatz, N. Wiser, E. Y. Tsymbal, and D. G. Pettifor, Phys. Rev. Lett. 85, 1314 (2000).
[11] J. Balogh, M. Csontos, D. Kaptás, and G. Mihály, Solid State Communications 126, 427 (2003).
[12] A. M. Goodman and S. J. Greaves, Y. Sonobe, H. Muraoka, and Y. Nakamura, J. Appl. Phys. 91, 8064 (2002).
[13] J. L. Dormann, in Advances in Chemical Physics, Vol. XCVIII, eds. I. Prigogine and Stuart A. Rice, 1997.
[14] K. Sumiyama, Vacuum 41, 1211 (1990)

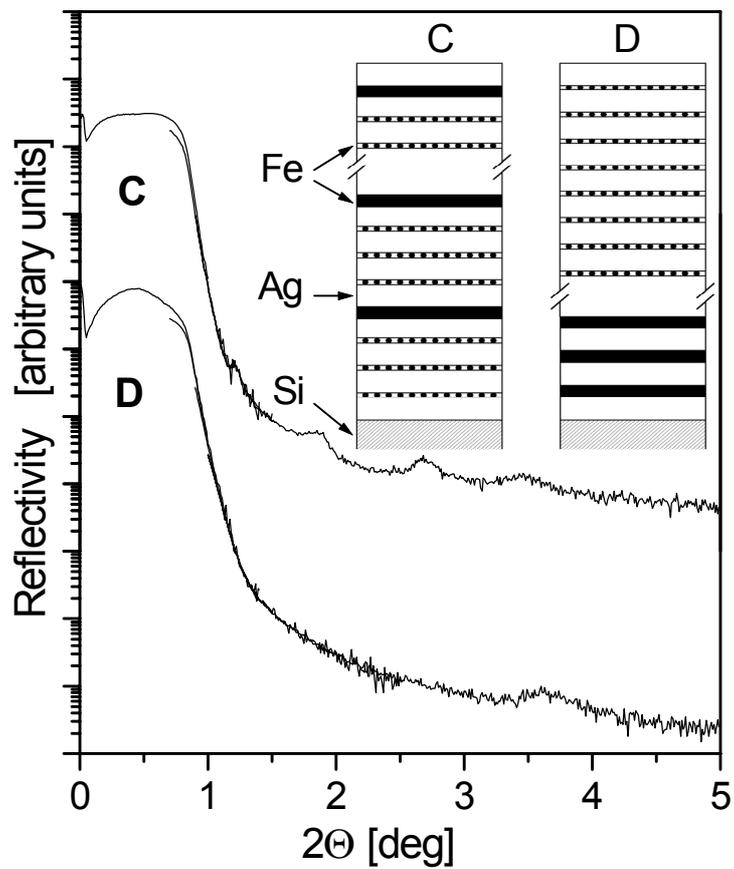

**Figure 1**

X-ray reflectivity of the interleaved (C) and the separated (D) sample. The inset shows a schematic representation of the sequence of the granular (dotted) and continuous (shaded) Fe layers in the samples.

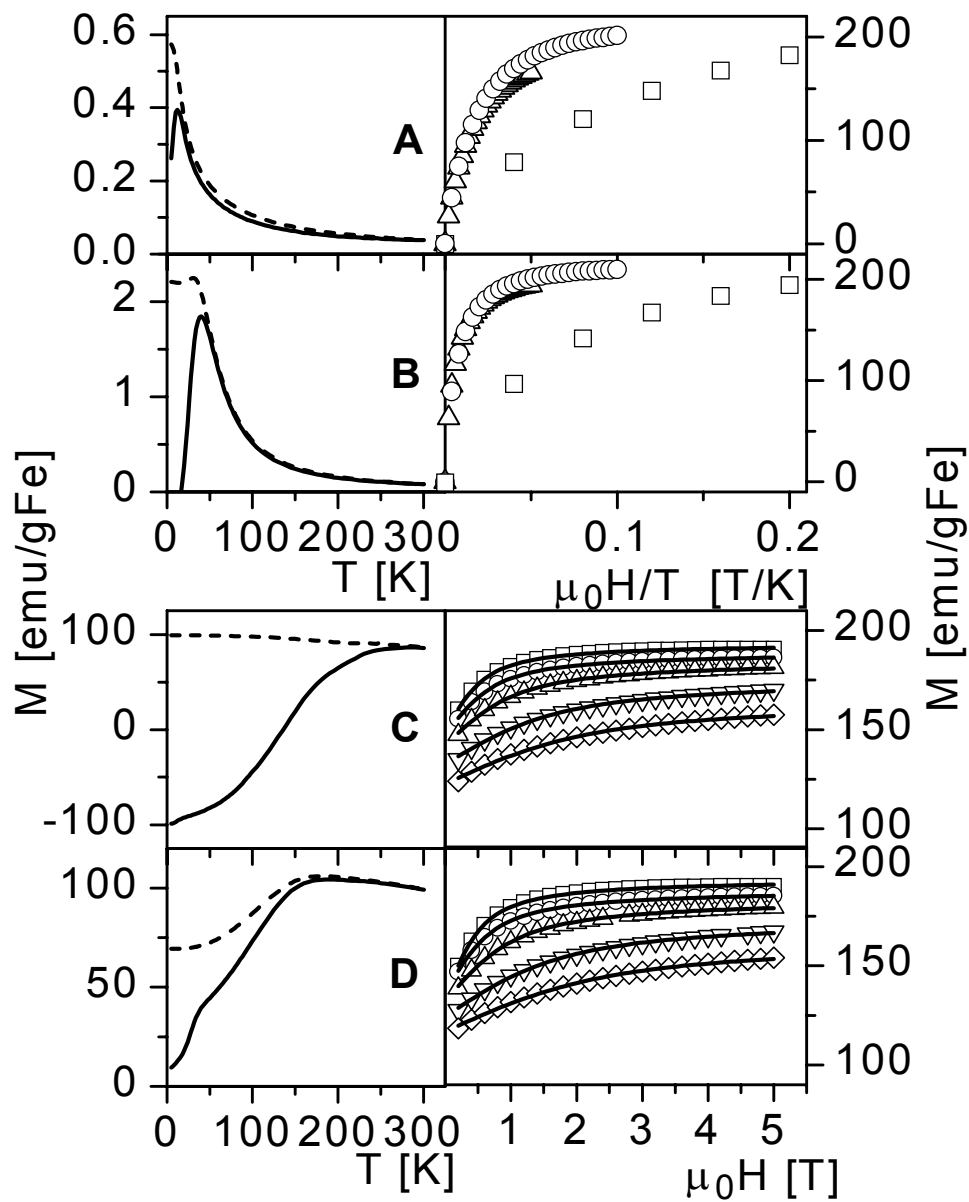

**Figure 2**

Left panels show the temperature dependence of the FC (dashed line) and ZFC (full line) magnetization measured in a 1 mT applied field for the granular multilayers (A and B) and for the interleaved (C) and separated (D) heterostructures. The right panels show the applied field dependence of the magnetization at 5 K (□), 50 K (○), 100 K (Δ), 200 K (∇) and 300 K (◊). For A and B the magnetization is plotted as a function of $\mu_0 H/T$. For C and D the curve fitted as the sum of a constant and a Langevin function is shown as well.

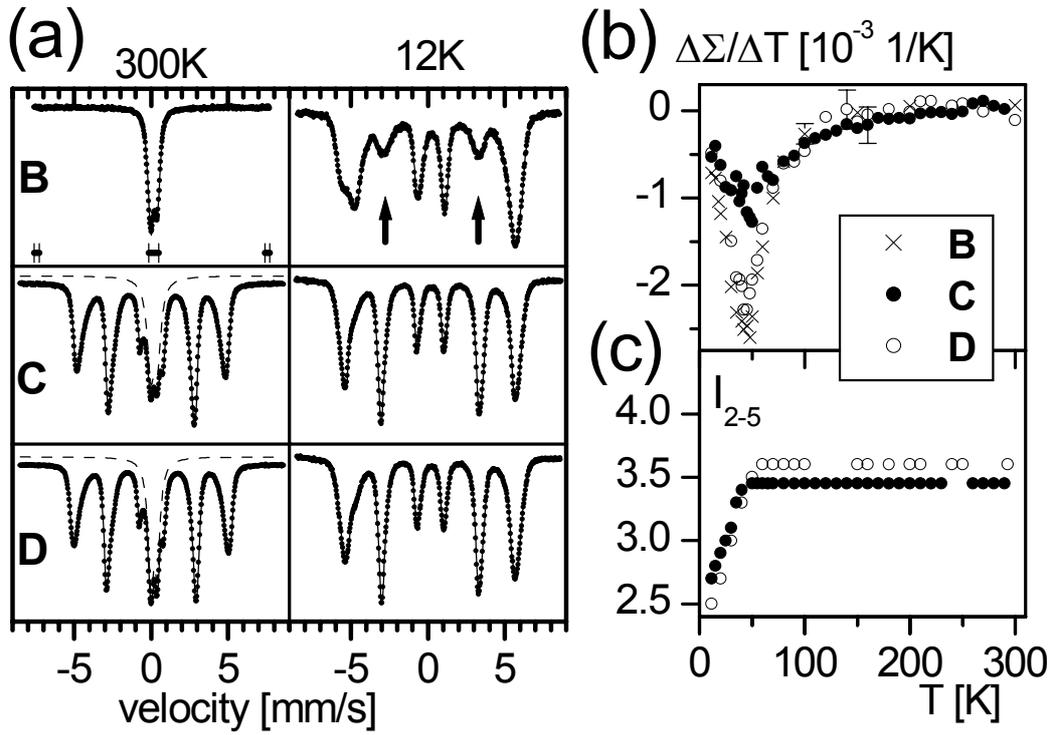

**Figure 3**

(a) Mössbauer spectra of samples B, C and D at 12 K and at 300 K. The velocity ranges indicated in the 300 K spectrum of sample B were used to calculate $\Sigma(T)$ (see text for details). The lines belonging to the $\Delta m=0$ magnetic transitions are marked by arrows in the 12 K spectrum of sample B (b) $\Delta\Sigma(T)/\Delta T$ for the granular (B), interleaved (C), and separated (D) samples. (c) Average intensity of the $\Delta m = 0$ transitions ($I_{2-5}$) of samples C and D.